\documentclass[twocolumn,aps,prb,showpacs,superscriptaddress]{revtex4}
\usepackage{amsmath,amssymb,bm}
\usepackage{graphicx,epsfig}
\usepackage{bm}

\newcommand{\be}{\begin{equation}}
\newcommand{\ee}{\end{equation}}
\newcommand{\nn}{\nonumber \\}
\newcommand{\ij}{\langle ij \rangle}

\newcommand{\ba}{\begin{eqnarray}}
\newcommand{\ea}{\end{eqnarray}}

\newcommand{\bpm}{\begin{pmatrix}}
\newcommand{\epm}{\end{pmatrix}}

\begin{document}
\setcounter{page}{1}
\title[]{Duality Argument for the Chiral-nematic Phase of Planar Spins}
\author{Jung Hoon \surname{Han}}
\affiliation{Department of Physics, BK21 Physics Research Division,
Sungkyunkwan University, Suwon 440-746, Korea}
\email{hanjh@skku.edu}

\begin{abstract}
A duality argument for the recently discovered chiral-nematic phase
of the XY model in a triangular lattice is presented. We show that a
new Ising variable naturally emerges in mapping the
antiferromagnetic $J_1 -J_2$ classical XY spin Hamiltonian onto an
appropriate Villain model on a triangular lattice. The new variable
is the chirality degree of freedom, which exists in addition to the
usual vortex variables, in the dual picture. Elementary excitations
and the associated phase transition of the Ising degrees of freedom
are discussed in some detail.
\end{abstract}



\maketitle

\section{Introduction}
A description of the excitations of a classical ferromagnet in a
two-dimensional lattice is based on two entities: spin waves, which
are small fluctuations of the spin orientation from their
ground-state direction, and vortices, which are topological
structures of spins carrying a non-zero winding number. Among the
two, the latter excitation is more effective at destroying the
coherence of spins and ultimately drives the phase transition
between a quasi-long-range ordered (QLRO) phase and a disordered,
paramagnetic phase \cite{KT}. A convenient mathematical description
of the phase transition in terms of the vortex variable is afforded
by the mapping first discussed by Villain (Villain mapping)
\cite{villain-map}.

There are instances, however, where vortices do not exhaust all the
relevant excitations for the phase transition. Another variable,
namely, chirality, often appears in models of classical spins with
frustration \cite{villain-chirality}. With its symmetry being
Ising-like, the chirality fluctuation can drive a second phase
transition, apart from the one driven by vortex proliferation, of
Ising universality class. A generalization of the Villain mapping
for spin models with frustration was given by Villain
\cite{villain-chirality}, where it was demonstrated that the
chirality emerges as a new, independent excitation mode of the
model.

As a most recent example of a chirality-driven phase transition, it
was shown in the context of a generalized antiferromagnetic XY spin
model on a triangular lattice that there can be a chirality-driven
phase transition taking place at temperatures well above the
magnetic transition temperature \cite{PONH}. The situation stands in
contrast with the case of standard antiferromagnetic XY model on the
same lattice, which only reveals a tiny separation between the
chirality order ($T_\chi)$ and the magnetic order ($T_\mathrm{KT}$)
temperatures \cite{literature}.

The aim of this paper is to place the observations made in Ref.
\onlinecite{PONH} in the context of the duality mapping of Villain
by identifying the proper Ising variables associated with the
chirality. While Villain's original paper \cite{villain-chirality}
considered a square lattice, we focus here on a triangular lattice,
as was the case for the numerical work of Ref. \onlinecite{PONH}. A
more elaborate discussion of Villain's original model appeared in
the work of Lee and Grinstein\cite{dhlee}, but their discussion was
for a square lattice, and the issue of the chirality did not arise
in their work.

\section{Duality Mapping}

The model studied in Ref. \onlinecite{PONH} reads

\be H = J_1 \sum_{\ij} \cos ( \varphi_{ij} )+ J_2 \sum_{\ij} \cos (
2\varphi_{ij} ) , \label{eq:our-model}\ee
where $\varphi_{ij}$ is the angle difference $\varphi_i -\varphi_j$
between nearest neighbors $\ij$. Focusing on a pair-wise interaction
$J_1 \cos [\varphi_{ij}] \!+\! J_2 \cos [2 \varphi_{ij}]$, the
minimum energy angle for a given $\ij$ bond is two-fold degenerate
at $\pi \pm \Delta$, $\Delta = \cos^{-1} (J_1 /4J_2)$, provided
$4J_2 > J_1$. For $4J_2<J_1$, a single minimum-energy angle is
achieved, and the duality mapping proceeds in the same manner as in
the ferromagnetic case.

\begin{figure}[t]
\centering
\includegraphics[scale=0.4]{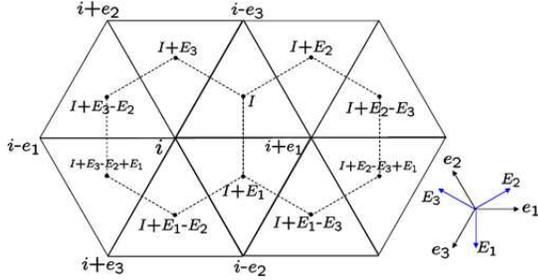}
\caption{Representation of the original triangular lattice (full
line) and the dual hexagonal lattice (dotted line) sites labeled by
lower case and upper case letters, respectively. Nearest-neighbor
vectors are denoted as $e_1$ ($E_1$) through $e_3$ ($E_3$) for the
triangular (hexagonal) lattice. }
\label{fig:dualattice}
\end{figure}

An appropriate Villain model for the interaction having a double
minima is given by the pairwise probability weight

\be P[\varphi_{ij}] = \sum_{n_{ij},t_{ij}} \exp\left( -{K\over 2} [
\varphi_{ij} \!-\! \pi \!-\! \Delta t_{ij}\!+\! 2\pi n_{ij} ]^2
\right). \ee
We introduce two integer link fields: $n_{ij}$, which run from
$-\infty$ to $\infty$, and $t_{ij}$, which takes on two values
$t_{ij}=\pm 1$ corresponding to two equivalent minima. Both fields
change sign under the interchange of the lattice indices,
$n_{ji}=-n_{ij}$, and $t_{ji} = -t_{ij}$. $K$ is related to the
inverse temperature $1/T$. Going through the standard Villain
mapping gives an alternative expression containing only the first
power of $\varphi_{ij}$:

\be P[\varphi_{ij}] = \sum_{l_{ij}, t_{ij}} \exp\left( -{1\over 2K}
l_{ij}^2 + i  l_{ij} (\varphi_{ij} \!-\!\pi \!-\! \Delta t_{ij})
\right), \ee
where $l_{ij}$ is another integer field running from $-\infty$ to
$\infty$. The partition function $Z$ is obtained as the product of
$P[\varphi_{ij}]$'s over all nearest-neighbor links:

\ba && Z = \prod_{\ij} P[\varphi_{ij}] =\sum_{\{l_{ij}, t_{ij}\}}
\prod_i \left(\int_{-\infty}^{\infty}  d\varphi_i \right) \nn
&&   \exp\left( -{1\over 2K} \sum_{\ij} l_{ij}^2 + i  \sum_{\ij}
l_{ij} (\varphi_{ij} \!-\!\pi \!-\! \Delta t_{ij}) \right) . \ea
The exponential factor in the above equation is evaluated for each
configuration $\{l_{ij}, t_{ij}\}$ and is summed over all possible
configurations to give $Z$. Integration over the angle $\phi_i$ can
be done first, which imposes current conservation at each site. In
turn, the constraint can be solved by introducing a dual integer
field $h_I$, which is connected to $l_{ij}$ by

\ba l_{i,i+e_1} &=& h_I - h_{I+E_1 }, \nn
l_{i,i-e_3} &=& h_{I+E_3}-h_{I}, \nn
l_{i,i+e_2} &=& h_{I - E_2+E_3 } - h_{I+E_3}, \nn
i_{i,i-e_1} &=& h_{I+ E_1 -E_2 + E_3} - h_{I-E_2 +E_3 }, \nn
i_{i,i+e_3} &= &h_{I+E_1 - E_2} - h_{I+E_1 - E_2 +E_3}, \nn
i_{i,i-e_2} &=& h_{I+E_1} - h_{I+E_1 -E_2}. \ea
See Fig. \ref{fig:dualattice} for the definitions of various labels.

Now, we have the partition function

\ba Z &= &\sum_{\{ h_I, t_{ij} \}} e^{-A}, \nn
A &= & {1\over 2K} \sum_{\langle IJ \rangle } (h_I \!-\! h_J )^2 + i
\sum_{\langle IJ \rangle } (h_I \!-\! h_J) (\pi \!+\! \Delta t_{ij})
, \nn\ea
where $A$ is given as the sum over the dual links $\langle
IJ\rangle$. The spin-wave contribution to $Z$ has been dropped. On
re-organizing the expression on the far right of $A$ for each $h_I$,
we get an equivalent expression

\be \sum_{\langle IJ \rangle } (h_I \!-\! h_J) (\pi \!+\! \Delta
t_{ij})=\sum_I h_I (3 \pi + \Delta c_I ) .\ee
Here, $c_I$ is the sum of the $t_{ij}$'s for each triangle centered
at $I$ and going in the \textit{counter-clockwise sense}. The
allowed values of $c_I$ are $c_I = \pm 3, \pm 1$. The partition
function is reduced to

\be Z \!=\! \sum_{\{ h_I, t_{ij} \}} \exp\left( -{1\over 2K}
\sum_{\langle IJ \rangle } (h_I \!-\! h_J )^2 \!-\! i \sum_{I} h_I
(\pi \!+\! \Delta c_I ) \right). \ee
At this point, we invoke the Poisson sum formula to re-write the
partition function as

\ba && Z = \sum_{\{m_I, t_{ij}\} } \prod_I
\left(\int_{-\infty}^\infty d\theta_I \right) \nn
&& \exp \left( -{1\over 2K}\sum_{\langle
IJ\rangle}(\theta_I-\theta_J )^2 + 2\pi i \sum_I M_I\theta_I
\right), \nn
&& M_I =m_I - {1\over 2} - {\Delta \over 2\pi} c_I  . \ea
Here, $m_I$ is another integer field from $-\infty$ to $\infty$
defined at the dual sites $I$, which will play the role of the
vorticity. We will integrate out $\theta_I$ to obtain the partition
function solely in terms of the vorticity $m_I$ and chirality $c_I$:

\be Z \sim  \sum_{\{m_I, t_{ij}\}}\exp \left(-2\pi^2 K \sum_{IJ}M_I
G_{IJ} M_J \right) . \ee
In the sum, $I$ and $J$ independently run over all dual-lattice
sites. The real-space Green's function, $G_{IJ}$, is given by

\be G_{IJ}=a^2 \int {{d^2 {\bf k}}\over {(2\pi)}^2}{ {e^{i{\bf
k}\cdot {\bf r}_{IJ}}} \over {4-2\cos(k_x a)-2\cos(k_y a)}}.\ee
It is convenient to use the regularized Green's function $G'_{IJ}$:
$G'_{IJ}=G_{IJ}-G_{{\bf 0}}$, so

\ba && Z=\sum_{\{m_I, t_{ij}\}}\exp\left( -2\pi^2 K G_{{\bf
0}}\left(\sum_I M_I \right)^2\right) \nn
&& ~~~~~\times\exp\left(-2\pi^2 K \sum_{IJ} M_I G'_{IJ} M_J \right).
\ea
Since $G_{{\bf 0}} \sim (1/2\pi)\ln (L/a)$, the allowed
configurations are those with total vorticity  $\sum_I M_I =0$. With
the help of the result at large distances

\be G'_{IJ}\sim -{1\over 2\pi} \ln {r_{IJ}\over a} -{1\over4},\ee
the partition function may be rewritten as

\ba Z\!=\!\sum_{\{m_I, t_{ij}\}} \exp\Bigl( -{\pi^2 \over 2} K
\sum_I M_I^2 + \pi K\sum_{I \neq J} M_I M_J \ln \left| {r_{IJ} \over
a} \right| \Bigr). \label{eq:vortex-action} \nn\ea
We have used the relation $(\sum_I M_I )^2 = 0 = \sum_I M_I^2 +
\sum_{I \neq J} M_I M_J $. Often the fugacity term is defined by
$y=\exp(-\pi^2 K /2)$. This term controls the number of (total)
vortices in the system. Equation (\ref{eq:vortex-action}) is the
desired action expressed solely in terms of the vorticity $m_I$ and
the chirality $c_I$ (or $t_{ij}$).

\begin{table}
\begin{tabular}{|c | c | c| c|}
\hline
$c_I$ & ${1\over 2} + {1\over 4} (1-\delta) c_I$ & $m_I$ & $M_I$ \\
\hline
3 & ${5\over 4} - {3\over 4}\delta$ & 1 & $-{1\over 4} + {3\over 4}\delta$ \\
\hline
-3 & $-{1\over 4} + {3\over 4}\delta$  & 0 & ${1\over 4} - {3\over 4}\delta$\\
\hline
1 & ${3\over 4} - {1\over 4}\delta$  & 1  &  ${1\over 4} + {1\over 4}\delta$ \\
\hline
-1 & ${1\over 4} + {1\over 4} \delta $  & 0 & $-{1\over 4} - {1\over 4}\delta$  \\
\hline
\end{tabular}
\caption{Local chirality $c_I$ and local vorticity $m_I$ for which
the net vorticity $M_I$ is least costly. }\label{table:vorticity}
\end{table}

In Ref. \onlinecite{PONH}, the large $J_2 /J_1$ region was shown to
have a chirality ordering transition taking place well above the
magnetic transition. In the $J_2 /J_1 \gg 1$ region, $\Delta =
\cos^{-1} (J_1 /4J_2)$ becomes close to $\pi/2$, and one can write
$\Delta = \pi (1-\delta)/2$, where $\delta$ is given by $\sin (\pi
\delta /2 ) = J_1 /4 J_2 $. The fugacity consideration requires that
$M_I^2$ be as small as possible at low temperatures, and in Table
\ref{table:vorticity}, the smallest net vorticity $M_I$ for each
chirality value $c_I$ is shown. Furthermore, $M_I^2$ for $c_I = \pm
3$ is less than $M_I^2$ when $c_I = \pm 1$,

\be M_I^2 (c_I = \pm 3) - M_I^2 (c_I = \pm 1) \approx -\delta /2 <
0. \ee
The global minimum of $\sum_I M_I^2$ is obtained if each triangle
carries the chirality $c_I = \pm 3$. In practice, this is possible
by arranging all the up triangles to carry $c_I = 3$ and all the
down triangles to carry $c_I = -3$, or vice versa. This is the
chirality-ordered phase at low temperature.

\begin{figure}[ht]
\centering
\includegraphics[scale=0.3]{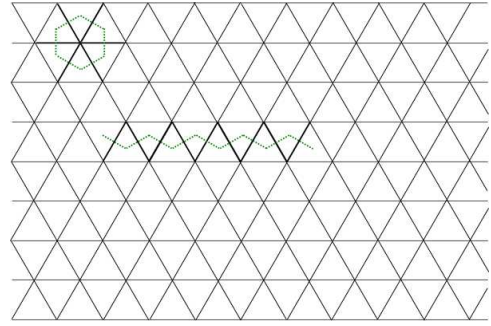}
\caption{The thin lines forming the triangular lattice have the
chirality $c_I = 3$ for upward triangles and $c_I =-3$ for downward
triangles. The thick lines have the local chirality $t_{ij}$
reversed from its ground state value, forming ``defect bonds." The
lines (green dotted) that intersect such defect bonds can form a
closed loop (the smallest such loop is a hexagon) or an open
segment. Each point of the line segment carries vorticity changes of
$\pm \delta$ if it is connected on both sides and $\pm (1/2 -\delta
/2)$ if it is connected on one side only. } \label{fig:defects}
\end{figure}

The cheapest excitation is the one with the least increase in
$M_I^2$. Such an excitation is achieved if for a given triangle, two
of the $t_{ij}$'s reverse their directions, and $c_I =3$ becomes
$c_I = -1$ or $c_I =-3$ becomes $c_I =1$. As one can see from Table
\ref{table:vorticity}, the corresponding change in the net vorticity
is from $M_I=-1/4+3\delta /4$ to $M_I =-1/4-\delta/4$, or $\Delta
M_I = -\delta$. Similarly $c_I =-3 \rightarrow 1$ incurs the change
$\Delta M_I = \delta$. Such excitations are called incommensurate
vortices \cite{incomm-vortex} to set them apart from integer changes
in the vorticity, also called commensurate vortices. At a
temperature $T_I$ that scales with $\delta$, we thus expect an Ising
transition caused by the proliferation of incommensurate vortices.
Imagine we draw a line segment from $I$ and $J$ that intersects the
$\ij$ bond with its $t_{ij}$ reversed. Then as one can see from Fig.
\ref{fig:defects}, the creation of incommensurate vortices is
achieved if the line segments close onto themselves, forming closed
loops. In the dual hexagonal lattice, the smallest such loop would
be a single hexagon. At the Ising transition, the size of the closed
loop diverges. In the numerical study of Eq. (\ref{eq:our-model}),
an Ising transition with $T_I \sim \delta$ was identified for the
large $J_2 /J_1$ region.

A flip of a single $t_{ij}$ in a triangle, on the other hand,
changes $c_I =3$ (-3) to $c_I=1$ (-1), and the net vorticity change
$\Delta M_I = \pm ( 1/2-\delta/2)$. Close to $\delta=0$, this is
creating an isolated half-integer vortex with a rather high energy
cost. To avoid the creation of a half-integer vortex, all triangles
must have $ \Delta c_I = \pm 4$, rather than $\Delta c_I = \pm 2$.
For an open line segment, if both ends of the line terminate at the
A or the B sublattice sites of the hexagonal lattice, the endpoints
are associated with half-integer vortices of the same sign. If one
end terminates on a sublattice different from that of the other end,
then one has created a pair of half-integer vortices of opposite
charges. It is then the proliferation of closed loops in the dual
hexagonal space that corresponds to the Ising transition of the
chirality. At exactly $\delta =0$ ($J_1 = 0$), the energy cost for
the flip of chirality by four units becomes zero, so already at zero
temperature, one has a disordered chirality phase. At small
$\delta$, a proliferation of the cluster of triangles with $\Delta
c_I = \pm 4$ or a proliferation of closed loops in the dual
hexagonal space occurs at a small finite temperature $T_I$. On the
other hand, once the length of the line becomes infinitely long, the
half-integer vortices becomes free, and the half-integer vortex
pairs unbind. This is the KT transition of half-integer vortices at
a higher temperature $T_\mathrm{KT}$.

\section{Summary}
To summarize, we have presented a duality argument for the
antiferromagnetic $J_1-J_2$ XY spin model on a triangular lattice.
In carrying out the Villain mapping of the original spin model, an
additional Ising degree of freedom emerges naturally. This new
degree of freedom corresponds to the spin chirality, and its
excitation is responsible for the chirality phase transition of
Ising universality. The natures of various possible Ising
excitations are clarified in the dual model.

\begin{acknowledgments}
The author thanks Dung-Hai Lee for the discussion that led to the
development of the ideas presented in this paper, and Jing-Hong
Park, Shigeki Onoda, and Naoto Nagaosa for an earlier collaboration.
This work was supported by the Korea Research Foundation grant
(KRF-2008-521-C00085, KRF-2008-314-C00101) and by a Korea Science
and Engineering Foundation (KOSEF) grant funded by the Korea
government(MEST) (No. R01-2008-000-20586-0).
\end{acknowledgments}

\newpage

\begin{thebibliography}{99}

\bibitem{KT} J. M. Kosterlitz and D. J. Thouless, J. Phys. C: Solid
State Phys. \textbf{6}, 1181 (1973).

\bibitem{villain-map} J. Villain, J. Physique \textbf{36}, 581 (1975).

\bibitem{villain-chirality} J. Villain, J. Phys. C: Solid State Phys.
\textbf{10}, 4793 (1977).


\bibitem{PONH} Jin-Hong Park, Shigeki Onoda, Naoto Nagaosa, and Jung
Hoon Han, Phys. Rev. Lett. \textbf{101}, 167202 (2008).

\bibitem{literature} P. Olsson, Phys. Rev. Lett. \textbf{75}, 2758 (1995);
S. Lee and K.-C. Lee, Phys. Rev. B \textbf{57}, 8472 (1998); S. E.
Korshunov, Phys. Rev. Lett. \textbf{88}, 167007 (2002); M.
Hasenbusch, A. Pelissetto, and E. Vicari, J. Stat. Mech.: Theory
Exp., P12002 (2005).

\bibitem{dhlee} D. H. Lee and G. Grinstein, Phys. Rev.
Lett. \textbf{55}, 541 (1985).

\bibitem{incomm-vortex} D. H. Lee, G. Grinstein, and J. Toner,
Phys. Rev. Lett. \textbf{56}, 2318 (1986).

\end{thebibliography}
\end{document}